\documentclass{elsart}[10pt]
\usepackage{epsfig}
\begin{document}
\begin{frontmatter}
\title{Exact broken-symmetry states and Hartree-Fock solutions for quantum dots at high magnetic fields}
\author{B. Szafran\thanksref{l1}\thanksref{l2}}, \author{F.M. Peeters\thanksref{l1}}, \author{S. Bednarek\thanksref{l2}}, and \author{J. Adamowski\thanksref{l2}}
\address[l1]{Departement Natuurkunde, Universiteit Antwerpen (Campus drie Eiken),
B-2610 Antwerpen, Belgium}
\address[l2]{Faculty of Physics and Nuclear Techniques, AGH University of Science and Technology, Krak\'ow, Poland}
\begin{abstract}
Wigner molecules formed at high magnetic fields in circular and
elliptic quantum dots are studied by exact diagonalization (ED)
and unrestricted Hartree-Fock (UHF) methods with multicenter basis
of displaced lowest Landau level wave functions. The broken
symmetry states with semi-classical charge density constructed
from superpositions of the ED solutions are compared to the UHF
results. UHF overlooks the dependence of the few-electron wave
function on the actual relative positions of electrons localized
in different charge puddles and partially compensates for this
neglect by an exaggerated separation of charge islands which are
more strongly localized than in the exact broken-symmetry states.
\end{abstract}
\end{frontmatter}
\newpage
At high magnetic fields the electron systems in circular quantum
dots form Wigner molecules \cite{rev,max} in the internal
structure of the system. Deformation \cite{MEPJD,SPBA} of the
circular symmetry allows the molecules to appear in the laboratory
frame only \cite{SPBA} when the classical counterpart
\cite{Bedanov} of the quantum system possesses a single
lowest-energy configuration. Otherwise the charge density at high
field is a superposition of equivalent semi-classical densities
and the quantum system undergoes symmetry transformations when the
magnetic field is increased \cite{SPBA}. These transformations are
associated with level crossings at which the ground state is
two-fold degenerate. Superposition of the states of the degenerate
levels allows \cite{SPBA} to extract the semi-classical
broken-symmetry charge density into the laboratory frame. On the
other hand the unrestricted Hartree-Fock (UHF) produces \cite{rev}
broken-symmetry states for Wigner molecules. In the infinite
magnetic field limit UHF gives exact \cite{SEPJD} results for the
total energy. At finite magnetic fields for which exact
broken-symmetry eigenstates exist the artifactal symmetry breaking
cannot be blamed to the inaccuracy of the  UHF \cite{SPBA}. In
this paper we look for the effects neglected by UHF comparing the
ED and UHF solutions for elliptical and circular dots.

We assume a spin-polarization of electrons at high magnetic field
$(0,0,B)$ oriented perpendicular to the quantum dot plane and use
the Landau gauge. In the ED, described in detail in Ref.
\cite{SPBA}, the single electron wave functions used for
construction of the Slater determinants are obtained via
diagonalization of the single electron Hamiltonian in the
multicenter basis \cite{K,Y,NHF1} of $M$ displaced lowest Landau
level wave functions
\begin{eqnarray}
\psi(\mathbf{r}) = & \sum_{k=1}^{M} c_k \exp\{-\alpha
[(x-x_k)^2+(y-y_k)^2]/4+ \nonumber \\ &
+ieB(x-x_k)(y+y_k)/2\hbar\},
\end{eqnarray}
where $\alpha$ is treated as variational parameter. In the present
UHF approach the one-electron orbitals (1) are optimized
self-consistently. We study up to $N=4$ electrons, use the
material data of GaAs and a basis of 12 centers $(x_k,y_k)$ put on
an ellipse with a size determined variationally. Basis (1) of
displaced lowest Landau level wave functions reproduces
\cite{SPBA} also higher Fock-Darwin bands. Contrary to previous
multicenter HF calculations \cite{SEPJD,NHF1} using a single
center per electron, the present HF approach produces results
which are exact in the UHF \cite{RG} sense.

Classical system of three electrons in an elliptical confinement
potential with $\hbar \omega_x=3$ meV and $\hbar \omega_y=4$ meV
possesses two equivalent lowest-energy configurations [cf. inset
of Fig. 1] and the quantum system undergoes parity transformations
\cite{SPBA} with the magnetic field [cf. Fig. 1]. Superposition
\cite{SPBA} of the two lowest-energy eigenstates
\begin{equation}
\Psi_{BS}=(\Psi_{even}+e^{i\phi}\Psi_{odd})/\sqrt{2}\end{equation}
yields a broken-symmetry (BS) charge density with a distinct
electron separation. Fig. 1 shows that in contrast to the exact
ground-state energy the UHF energy estimate is a smooth function
of the magnetic field.

The charge densities of considered states are shown in Figs. 2(a)
and 2(b) for two magnetic field values corresponding to the
even-odd energy crossing presented in Fig. 1. The phase $\phi$ in
the BS state is chosen such that the electrons are localized at
the classical Wigner molecule positions. Notice that in the UHF
the separation of electrons is more pronounced than in the exact
BS states. Fig. 2(c) shows the pair correlation function (PCF)
\cite{max} for the UHF and the exact BS state corresponding to the
charge density of Fig. 2(a) with the position of one of the
electrons fixed at two different locations: in the center and on
the edge of the central charge puddle. In contrast to the exact BS
state in the UHF wave function the two electrons are insensitive
to the actual position of the third electron in its charge puddle.
This is a consequence of the single-determinantal form of the UHF
wave function, and can be easily explained for two electrons. In
the spin-polarized two electron Wigner molecule the UHF spatial
wave function is given by $\Psi_\alpha({\bf r}_1 )\Psi_\beta({\bf
r}_2)-\Psi_\beta({\bf r}_1)\Psi_\alpha({\bf r}_2)$, where
$|\Psi_\alpha|^2$ and $|\Psi_\beta|^2$ are the charge densities of
separate charge puddles $\alpha$ and $\beta$. Wave functions
$\Psi_\alpha$ and $\Psi_\beta$ are orthogonal due to the vanishing
overlap between the puddles. The calculation of the PCF gives (up
to a constant) PCF$({\bf r}_a,{\bf r}_b)= |\Psi_\alpha({\bf
r}_a)|^2|\Psi_\beta({\bf r}_b)|^2+|\Psi_\alpha({\bf
r}_b)|^2|\Psi_\beta({\bf r}_a)|^2$. For ${\bf r}_b$ inside puddle
$\beta$ the second term of the sum vanishes and the remaining one
signifies that the probability of finding an electron in point
${\bf r}_a$ inside puddle $\alpha$ is independent of the actual
position of the second electron in puddle $\beta$.

The UHF self-consistency is reached only in one of the two
classical orientations [cf. inset of Fig. 1] in which the UHF
energy is minimal. On the other hand, the exact BS states can be
oriented under an arbitrary angle [cf. Fig. 2(d)] by modifying the
phase $\phi$ in Eq. (2). Moreover, since the BS state is
constructed with states of opposite parities all the plots in Fig.
2(d) correspond to the same value of the kinetic, potential and
electron-electron interaction energies equal to the arithmetic
average of the expectation values for $\Psi_{odd}$ and
$\Psi_{even}$ states.

Fig. 3 shows the two lowest energy levels and the UHF total energy
calculated with respect to the lowest Landau level for the
elliptical dot with $\hbar \omega_x=3$ meV and $\hbar \omega_y=4$
meV. For these values the classical counterpart of the
four-electron system is unique and conform with the symmetry of
the confinement potential, so that the Wigner crystallization is
visible in the exact quantum ground-state for an arbitrary
magnetic field after the MDD decay. In this case the MDD decay is
a continuous process and appears at the anticrossing around 6 T.

The inset to Fig. 3 shows the charge densities calculated with the
UHF and ED methods. In ED the charge density in
 between the charge maxima takes on larger values than in UHF in
 which the separation of electrons is more distinct like for
 $N=3$ [cf. Fig. 2(a) and (b)].

The average radius of the charge puddle as obtained for the
two-electron ground state in the circular quantum dot ($\hbar
\omega_x=\hbar\omega_y=3$ meV) is displayed in Fig. 4. The ED and
UHF values are similar below the MDD breakdown ($B<5.6$ T). The
exact value present discontinuities at the angular momentum
transitions. After the MDD decay the UHF value is close to the
average around which ED and BS results oscillate, but at higher
fields it becomes an upper bound for these oscillations. The inset
of Fig. 4 shows the exact BS and UHF charge density for $B=11.8$
and 28.9 T. BS charge densities for the two values of the magnetic
field have been obtained from superpositions of the degenerate
states with angular momenta -5,-7 and -9,-11 (in $\hbar$ units),
respectively. The charge maxima in the exact BS solutions are less
strongly separated. For $B=28.9$ T the charge density islands of
the exact BS have crescent shape while the UHF charge puddles are
more oval and the distance between them is larger.

The difference in shape of the separated charge islands in
circular dots is largest for $N=2$. For larger $N$ the charge
puddles in the exact solutions are less spectacularly spread. Fig.
5 shows the comparison of the exact BS and UHF charge densities
for $N=3$. The BS plots for $N=3$ correspond to degeneracy of
states with angular momentum $-9\hbar$, $-12\hbar$ for $B=7.5$ T
and $-12\hbar, -15\hbar$ for $B=15$ T.

In summary, we have presented a comparison of the UHF and the ED
results for the charge density of Wigner molecules in circular and
elliptical dots. For the comparison we have used the
broken-symmetry states obtained from the superposition of the
exact eigenstates. We have found that the UHF exaggerates the
separation of the electron charge densities in the laboratory
frame. In this way the UHF method partially compensates for the
overlooked correlations related to the reaction of electrons on
their actual position in the separated charge density islands.
This reaction is of smaller importance for larger magnetic fields,
for which the charge density islands shrink to points, which
explains the vanishing of the UHF energy overestimation in the
infinite magnetic field \cite{SEPJD}. Due to the exaggerated
electron separation the charge islands forming Wigner molecules
calculated in the UHF shrink with the magnetic field faster than
in the exact broken-symmetry states.

{\bf Acknowledgments} This paper was supported in part by the
Polish Ministry of Scientific Research and Information Technology
in the framework of the solicited grant PBZ-MIN-008/P03/2003, the
Flemish Science Foundation (FWO-Vl), the Belgian Science Policy,
and the University of Antwerpen (VIS and GOA). One of us (BS) is
supported by the Foundation for Polish Science (FNP).

\begin{figure}[htbp]{\epsfxsize=55mm
                \epsfbox[47 110 571 630]{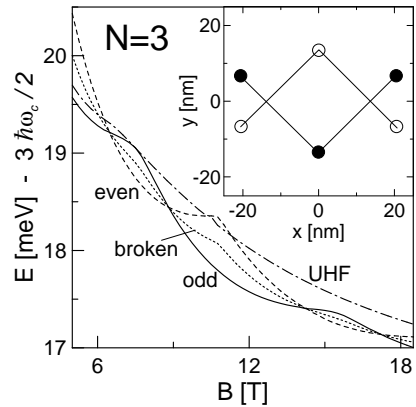}}\newline
\caption{Energy of the lowest even (dashed curve) and odd parity
(solid curve) levels, the broken symmetry state (dotted curve) and
UHF energy calculated for $N=3$, $\hbar \omega_x=3$ meV and $\hbar
\omega_y=4$ meV. Inset shows the two equivalent classical
configurations.}
\end{figure}

%\begin{figure*}[htbp]{\epsfxsize=120mm
%                \epsfbox[151 626 422 778]{fig2.eps}}\newline
\begin{figure}[htbp]{\epsfxsize=50mm
                \epsfbox[75 496 218 810]{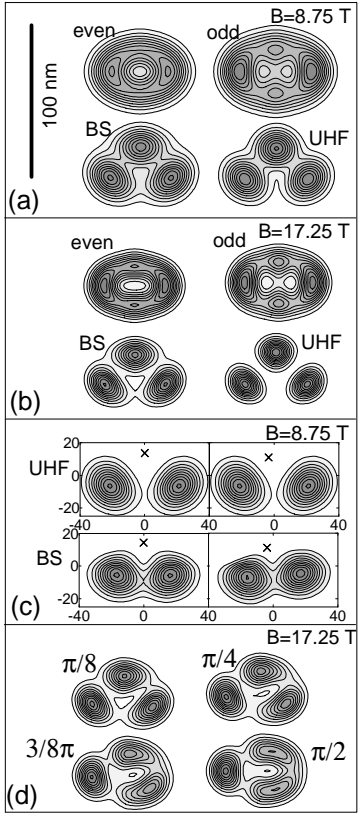}}\newline
                \caption{
(a) The charge densities for the even, the odd, the
broken-symmetry (BS) state and the UHF for $N=3$, $\hbar
\omega_x=3$ meV, $\hbar \omega_y=$ 4 meV at $B=8.75$ T. (b) Same
as (a) at $B=17.25$ T. (c) Pair correlation function plots for UHF
and BS state for an electron fixed at (0,13.4 nm) [left panel] and
(-3 nm,10.4 nm) [right panel] marked by crosses. (d) BS states for
given shifts of phase $\phi$ [Eq. (2)] with respect to the BS plot
in (b). Plots (a), (b) and (d) have the same scale given by the
length bar in (a).
 }
 \end{figure}
%\end{figure*}
\begin{figure}[htbp]{\epsfxsize=60mm
                \epsfbox[33 80 571 603]{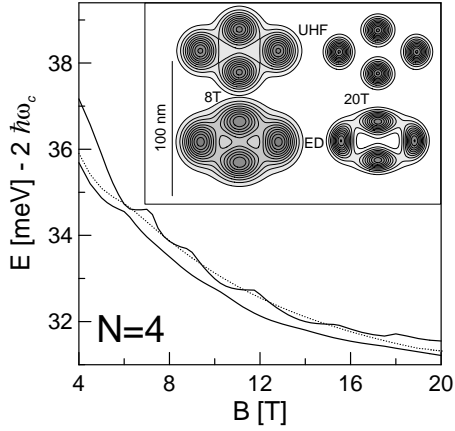}}\newline
                \caption{
Lowest energy levels for $N=4$, $\hbar \omega_x=3$ meV and $\hbar
\omega_y=4$ meV (solid lines) and total UHF energy (dotted) line.
Inset: Charge density calculated in the UHF (two upper plots) and
the ED (two lower plots) for $B=8$ T (left side) and $B=20$ T
(right side).
 }
 \end{figure}
\begin{figure}[htbp]{\epsfxsize=60mm
                \epsfbox[40 110 584 635]{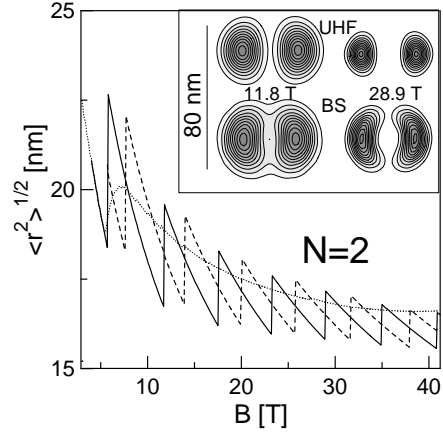}}\newline
                \caption{
Average radius of the charge puddle ($<r^2>^{1/2}$) for two
electrons in a circular quantum dot ($\hbar
\omega_x=\hbar\omega_y=3$ meV). Dotted, dashed and solid lines
correspond to UHF, BS, and ED results. Inset: charge density
obtained in the UHF (two upper plots) and in the exact
broken-symmetry state (two lower plots) for B=11.8 T (left side)
and 28.9 T (right side).}
 \end{figure}

\begin{figure}[htbp]{\epsfxsize=45mm
                \epsfbox[58 209 527 627]{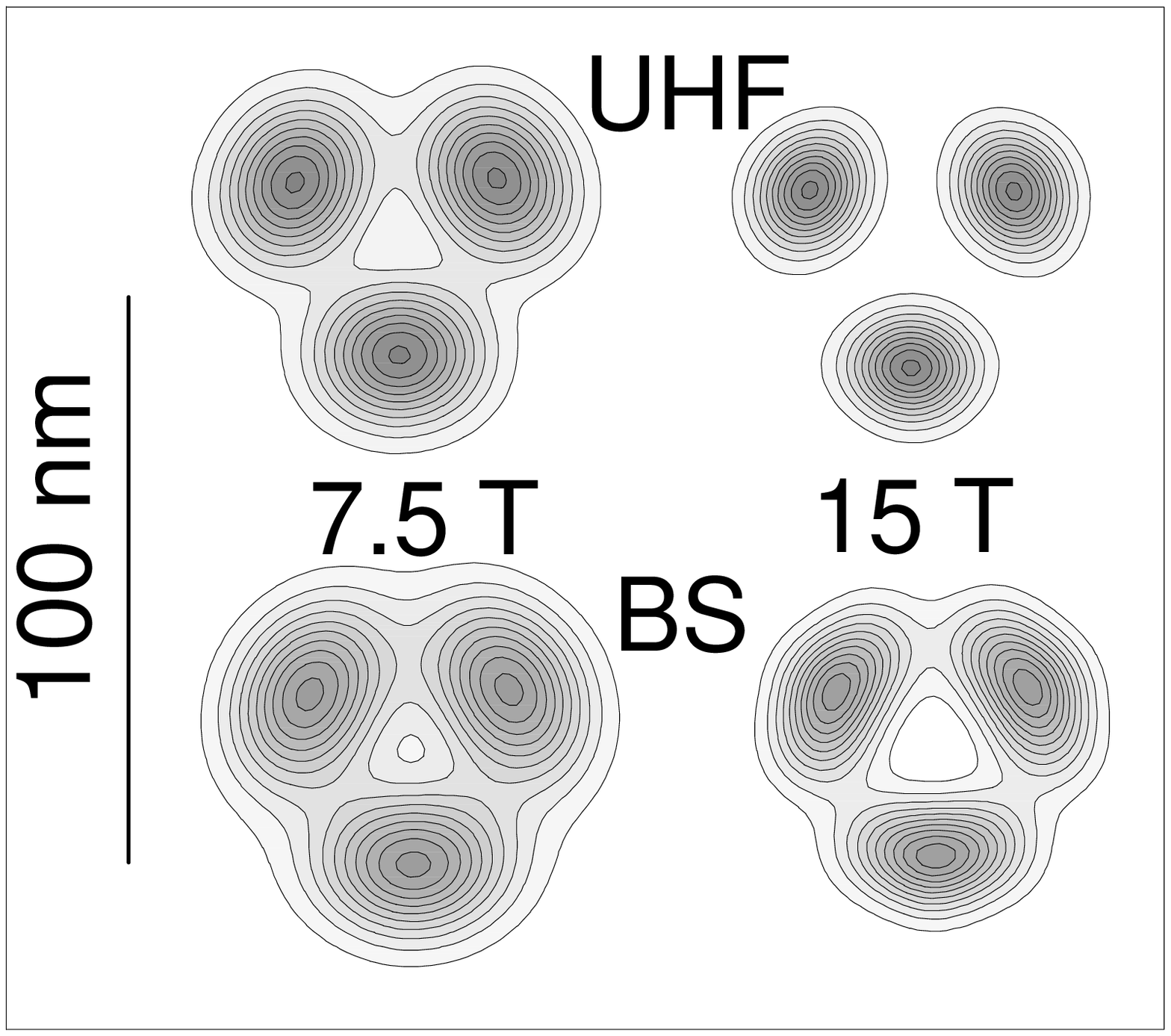}}\newline
                \caption{
Charge density for $N=3$ electrons in a circular quantum dot
($\hbar \omega_x=\hbar\omega_y=3$ meV) calculated with UHF (upper
plots) and for exact BS states (lower plots) for $B=7.5$ T (left
plots) and $B=15$ T (right plots). }
 \end{figure}


\begin{thebibliography}{00}

\bibitem{rev} S.M. Reimann and M. Manninen, Rev. Mod. Phys. 74 (2002)
1283.
\bibitem{max} P.A. Maksym, H. Immamura, G.P. Mallon, and H. Aoki, J. Phys. Condens. Matter 12 (2000)
R299.
\bibitem{MEPJD} M. Manninen, M. Koskinen, S.M. Reimann, and B.
Mottelson, Eur. Phys. J. D. 16 (2001) 381.
\bibitem{SPBA} B. Szafran, F.M. Peeters, S. Bednarek, and J.
Adamowski, Phys. Rev. B 69 (2004) 125344.
\bibitem{Bedanov} V.M. Bedanov and F.M. Peeters, Phys. Rev. B
49 (1994) 2667.
\bibitem{SEPJD} B. Szafran, S. Bednarek, J. Adamowski, M. Tavernier, E. Anisimovas, and F.M. Peeters,
Eur. Phys. J. D 28 (2004) 373.
\bibitem{K} J. Kainz, S.A. Mikhailov, A. Wensauer, and U. R\"ossler, Phys. Rev. B 65 (2002)
115305.
\bibitem{Y} C. Yannouleas and U. Landman, Phys. Rev. B {\bf 66} (2002)
115315.
\bibitem{NHF1} B. Szafran, S. Bednarek, and J. Adamowski, Phys. Rev. B
67 (2003) 045311; J. Phys.: Cond. Matter 15 (2003) 4189.
\bibitem{RG}  B. Reusch and H. Grabert,
Phys. Rev. B 68 (2003) 045309.

\end{thebibliography}
\end{document}